\documentclass[a4paper,showpacs,prb,twocolumn]{revtex4}
\usepackage{amsfonts}
\usepackage{amsmath}
\usepackage{amssymb}
\usepackage{graphicx}
\usepackage{verbatim}
\usepackage{color}

\begin{document}

\title{Interacting electrons in graphene nanoribbons in the lowest Landau level}
\author{A. A. Shylau}
\email{artsem.shylau@itn.liu.se}
\author{I. V. Zozoulenko}
\email{igor.zozoulenko@itn.liu.se} \affiliation{Solid State
Electronics, ITN, Link\"{o}ping University, 601 74, Norrk\"{o}ping,
Sweden}

\begin{abstract}
We study the effect of electron-electron interaction and spin on
electronic and transport properties of gated graphene nanoribbons (GNRs) in a
perpendicular magnetic field in the regime of the lowest Landau level (LL).
The electron-electron interaction is taken into account using the Hartree
and Hubbard approximations, and the conductance of GNRs is calculated on
the basis of the recursive Greens function technique within the Landauer
formalism. We demonstrate that, in comparison to the one-electron picture,
electron-electron interaction leads to the drastic changes in the dispersion
relation and structure of propagating states in the regime of the lowest LL
showing a formation of the compressible strip and opening of additional
conductive channels in the middle of the ribbon. We show that the latter are
very sensitive to disorder and get scattered even if the concentration of
disorder is moderate. In contrast, the edge states transport is very robust
and can not be suppressed even in the presence of a strong spin-flipping.
\end{abstract}

\date{\today }
\pacs{72.80.Vp, 73.22.Pr, 73.43.Cd, 73.43.Qt}
\maketitle

\section{Introduction}

Due to Dirac spectrum graphene possesses a number of unique
electronic and transport properties\cite{Castro_Neto_review}. In a
magnetic field $B$ perpendicular to a graphene layer the spectrum is
modified into a series of Landau levels (LL) with the energies
$E=\pm \hbar \omega _{c}\sqrt{N}$, where the cyclotron frequency
$\omega _{c}=v_{F}\sqrt{2eB/\hbar }$, $v_{F}\approx 10^{8}$ cm/s is
the Fermi velocity and $N=0,\pm 1,\pm 2,\ldots .$\cite {Gusynin} One
of interesting features of this spectrum is a presence of 0'th LL
with $E=0$, which is equally shared between electrons and holes. In
contrast to conventional two-dimensional electron gas systems the
position of this level does not depend on a value of the magnetic
field. An experimental manifestation of these unusual series is the
anomalous quantum Hall effect with the Hall conductivity given by
$\sigma_{xy}=\frac{4e^{2}}{h} (N+1/2)$, where the factor $4$ comes
from the spin and valley degeneracy\cite {Novoselov,Kim}. In a
strong magnetic field the four-fold degeneracy of the lowest (i.e.
0'th) LL is lifted leading to the insulating behavior characterized
by the presence of additional plateaux in the Hall conductivity
$\sigma_{xy}$ accompanied by the increase of the longitudinal
resistivity $\rho_{xx}$.\cite{Checkelsky,Zhang,Zaliznyak} The large
energy gaps identified experimentally suggests that electro-electron
interactions along with the Zeeman spin-splitting play an important
role in understanding of this phenomena.

Most of the experiments reported so far are performed in the
four-terminal geometry which allows a direct measurement of the
components of the resistivity and conductivity tensors\cite
{Checkelsky,Zhang,Zaliznyak,Giesbers}. However in some cases the
only possible experimental setup is the two-terminal measurements,
even though the results obtained in this geometry are not as
straightforward to interpret as the four-terminal
ones\cite{Williams}. A difference between the longitudinal and the
Hall resistances is not clearly defined in such the measurements.
The unexpected consequence of analysis of the two-terminal geometry
is that a measured two-terminal resistance between the right and
left contacts, $R_{2t},$ corresponds, in the four-terminal setup, to
the Hall resistance $R_{H}$ rather than to the longitudinal one
$R_{L}$. This stems from the fact that the voltage drop between the
sample edges, $V$, equals to the difference of the chemical
potentials in the contacts $eV=\mu _{R}-\mu
_{L}$.\cite{Kramer,Goerbig_Lecture}

An explanation of the quantum Hall effect is often based on the
picture of edge states. In a strong enough magnetic field the right
and left propagating states are localized at different edges of the
system due to the Lorentz force. It leads to exponentially small
overlap between counterpropagating states which, in turn, greatly
suppresses a possibility of backscattering and is eventually
manifested by developing pronounced plateaux in the conductance. In
graphene another transport regime can be realized\cite{Abanin}.
Coupling of spin to magnetic field due to the Zeeman effect leads to
a splitting of the lowest LLs into two sublevels. For the neutral
graphene the chemical potential coincides with the zero energy and
lies in the center of the spin-gap, see Fig. \ref{fig:BS} (d) below.
At a given Fermi energy there are two forward propagating states of
opposite spins located near different edges. The most prominent
feature of this transport regime is that forward and backward
propagating states\ of the different spins are not spatially
separated. If a scattering event leads to a spin flipping, the
overlap between the counterpropagating states of different spins
might be sufficient to cause the backscattering. This backscattering
of the spin-polarized states is used for an explanation of
the non-zero behavior of $\rho _{xx}$ near the Dirac point\cite%
{Abanin_Dissipative}.

It should be stressed that the structure of edge states derived
within a one-electron approximation and shown on Fig.(\ref{fig:BS})
(d) corresponds to the case of neutral graphene when the electron
density is zero. It is often used for understanding and explanation
of a variety of experiments on magnetotransport\cite{Ong}. The
conditions for existence of the edge states in GNRs was discussed in
details in Ref.[\cite{Gusynin_edge}]. Also, self-consistent
Hartee-Fock calculations were performed for GNR in magnetic
field\cite{Jung}. It was shown than the favored ground states in the
neutral graphene are charge- or spin-density waves. Neither of these
orderings support the edge states. However, in experiments one
usually needs to tune a charge concentration by means of the gate
voltage. It was shown before that electron-electron interaction
between induced charges can dramatically change the energy
dispersion and the structure of propagating
states\cite{Silvestrov,Shylau_Cap,Shylau_Magnet}. The magnetosubband
structure and the character of propagating states in the lowest LL
for the case of nonzero induced charge density remain largely
unexplored and represent the main focus of the present study. We
investigate the effect of the electron-electron interaction, spin
and global gate electrostatics on magnetotransport properties of GNR
in the vicinity of the charge neutrality point. Based on the
calculated self-consistent magnetosubband structure we compute and
analyze the two-terminal conductance of realistic GNRs focussing on
the effect of disorder, surface warping and spin flipping on the
robustness of the edge state transmission.

\section{Model}

\begin{figure}[tbh]
\includegraphics[keepaspectratio,width=0.8\columnwidth]{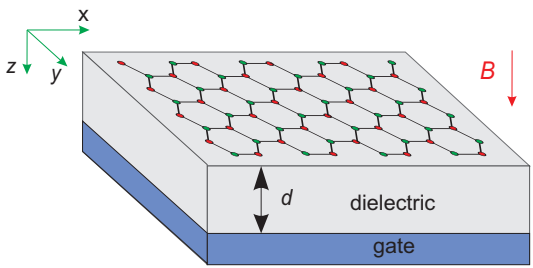}
\caption{(Color online). A schematic layout illustrating a system consisting
of a graphene nanoribbon of the width $w=20$ nm which is located on an
insulating substrate at the distance $d=300$ nm apart from a metallic back
gate. The whole system is subjected to a perpendicular magnetic field $B=150$
T.}
\label{fig:layout}
\end{figure}

We consider a system consisting of a GNR of the width $w=20$ nm and the
length $l=85$ nm, located on an insulating substrate $(\epsilon _{r}=3.9)$
and connected to ideal graphene leads of the same width, see Fig. \ref%
{fig:layout}. A metallic back gate is used to tune the Fermi energy and the
charge concentration in the GNR. The whole system is subjected to a uniform
magnetic field, $B=150$ T, perpendicular to the graphene plane. Such the
high value of the magnetic field allows us to use relatively narrow GNR (in
order to reduce computational time) and choose the ratio $w/l_{B}\approx 9.5$
in an accordance with typical experiments (with $l_{B}=\sqrt{\hbar /eB}$
being the magnetic length). In order to model magnetotrasport in a GNR we
use the tight-binding Hubbard Hamiltonian in the mean-field approximation $%
H=H_{\uparrow }+H_{\downarrow }$ which is shown to describe carbon electron
systems in good agreement with first-principles calculations\cite{Palacios},
\begin{eqnarray}
&&H_{\sigma }=-\sum_{\mathbf{r},\Delta }t_{\mathbf{r},\mathbf{r}+\Delta }a_{%
\mathbf{r}\sigma }^{+}a_{\mathbf{r}+\Delta ,\sigma }+g\mu _{b}B\sigma \sum_{%
\mathbf{r}}a_{\mathbf{r}\sigma }^{+}a_{\mathbf{r}\sigma }  \label{H} \\
&&+\sum_{\mathbf{r}}V_{H}(\mathbf{r})a_{\mathbf{r}\sigma }^{+}a_{\mathbf{r}%
\sigma }+U\sum_{\mathbf{r}}\left( \langle n_{\mathbf{r}\sigma ^{\prime
}}\rangle -\frac{1}{2}\right) a_{\mathbf{r}\sigma }^{+}a_{\mathbf{r}\sigma },
\notag
\end{eqnarray}%
where $\sigma $, $\sigma ^{\prime }$ describe two opposite spin states $%
\uparrow $, $\downarrow $; the summation runs over all sites of the graphene
lattice, $\Delta $ includes the nearest neighbors only, $t_{\mathbf{r},%
\mathbf{r}+\Delta }=t_{0}\exp (i2\pi \phi _{\mathbf{r},\mathbf{r}+\Delta
}/\phi _{0})$ with $t_{0}=2.77$ eV, $\phi _{0}=h/e$ being the magnetic flux
quantum which is calculated using the Pierel's substitution
\begin{equation}
\phi _{\mathbf{r},\mathbf{r}+\Delta }=\int_{\mathbf{r}}^{\mathbf{r}+\Delta }%
\mathbf{A}\cdot d\mathbf{l},  \label{phi}
\end{equation}%
with $\mathbf{A}$ being the vector potential. (In our calculations we use
the Landau gauge, $\mathbf{A}=(-By,0)$). The first term in the Hamiltonian (%
\ref{H}) describes hopping of electrons through the graphene
lattice. The second term describes the spin-splitting in the
magnetic field $B$ due to the Zeeman effect. Both these terms
correspond to the non-interacting part of the Hamiltonian. The
interaction between electrons of the opposite spins
is described by the Hubbard term with $U=1.33t_{0}$\cite{Yazyev, Jung}, and $%
V(\mathbf{r})$ is a Hartree term describing the long-range Coulomb
interaction between induced charges in the GNR\cite{Shylau_Cap,FR},
\begin{equation}
V_{H}(\mathbf{r})=\frac{e^{2}}{4\pi \varepsilon _{0}\varepsilon _{r}}\sum_{%
\mathbf{r}^{^{\prime }}\neq \mathbf{r}}n(\mathbf{r}^{^{\prime }})\left(
\frac{1}{|\mathbf{r}-\mathbf{r}^{^{\prime }}|}-\frac{1}{\sqrt{|\mathbf{r}-%
\mathbf{r}^{^{\prime }}|^{2}+4d^{2}}}\right) ,  \label{Vr}
\end{equation}%
where $d$ is the distance between the GNR and the gate, and the second term
describes a contribution from the mirror charges. In order to calculate the
electron number at site $\mathbf{r}$, Eq.(\ref{H}) is solved using the
recursive Green's function technique\cite{Xu}
\begin{equation}
\langle n_{\mathbf{r}\sigma }\rangle =-\frac{1}{\pi }\int_{-\infty }^{\infty
}\Im \lbrack G_{\sigma }(\mathbf{r},\mathbf{r},E)]f_{FD}(E,E_{F})dE,
\label{n}
\end{equation}%
where $G_{\sigma }(\mathbf{r},\mathbf{r},E)$ is the Green function in the
real space representation of an electron of the spin $\sigma $ residing on
the site $\mathbf{r}$ and $E_{F}=eV_{g}$ is the Fermi energy which position
is adjusted by the gate voltage. Equations (\ref{H})-(\ref{n}) are solved
self-consistently in order to compute the band structure, the charge density
and the potential distribution. For a given potential distribution we
compute the conductance using the Landauer formula
\begin{equation}
G(E_{F},B)=G_{0}\int T(E,B)\left[ -\frac{\partial f_{FD}(E-E_{F})}{\partial E%
}\right] dE,  \label{G}
\end{equation}%
where $T(E,B)$ is a total transmission coefficient, $G_{0}=e^{2}/h$ is the
conductance quantum, and $f_{FD}$ is the Fermi-Dirac distribution.

\section{Results and discussion}

\begin{figure*}[tbh]
\includegraphics[keepaspectratio,width=1.5\columnwidth]{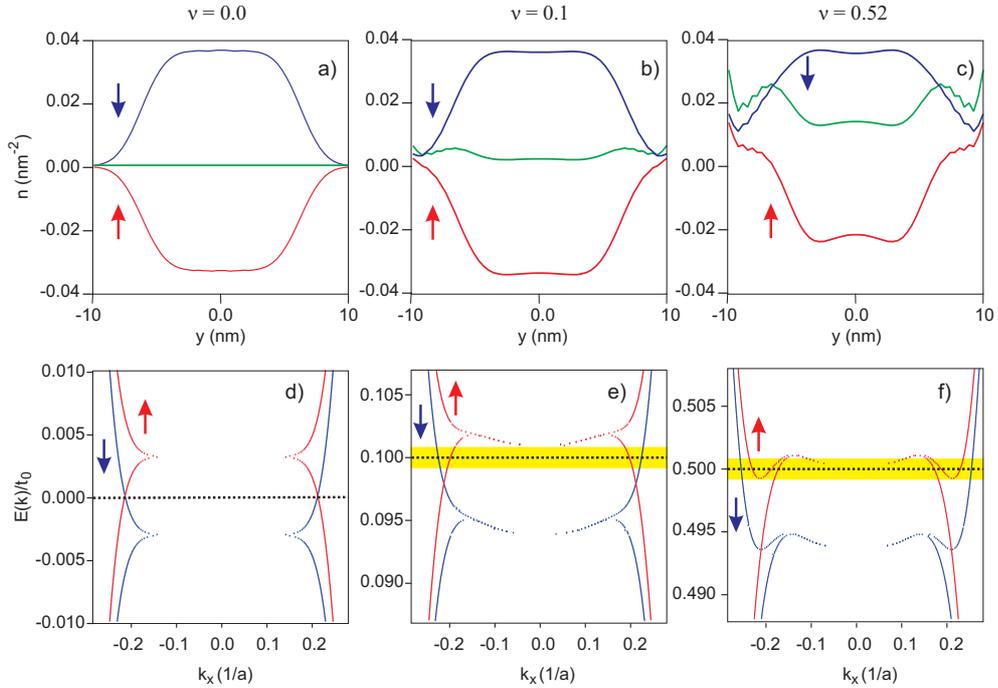}
\caption{(Color online). (a)-(c) Charge density distributions, and
(d)-(f) band structure of a graphene nanoribbon in a perpendicular
magnetic field $B=150$T. (a),(d) one-electron approximation,
$eV_{g}/t_0=0$; (b),(e) self-consistent calculations for
$eV_{g}/t_0=0.1$ and (c),(f) $eV_{g}/t_0=0.5$. Red and blue curves
marked by $\uparrow $ and $\downarrow $ correspond to the charge
densities of spin-up and spin-down electrons respectively,
$n_{\uparrow }(y),n_{\downarrow }(y)$. Charge densities are averaged
over three successive sites. Green curves correspond to the total
density distribution $n(y)=n_{\uparrow }(y)+n_{\downarrow }(y)$.
Dashed lines define the Fermi energy position. Yellow fields
correspond to the energy interval $[-2\protect\pi
k_{B}T,2\protect\pi k_{B}T]$; temperature $T=4.2$ K.} \label{fig:BS}
\end{figure*}

Figure \ref{fig:BS} shows the charge density distribution (a)-(c)
and the magnetosubband structure (d)-(f) of a GNR for several gate
voltages, $eV_{g}/t_0=0$,
0.1, and 0.5 respectively in the regime of the lowest LL, $|\nu |<1$ ($%
\nu =nh/eB$ being the filling factor and $n$ is the average charge
concentration). An application of the gate voltage induces extra charges in
the system, Fig. \ref{fig:BS} (b),(c), which leads to an asymmetry between
the electron and hole branches in the energy spectrum, see Fig. \ref{fig:BS}
(e),(f). For small gate voltages, see Fig. \ref{fig:BS} (e), the lowest
unoccupied spin-up branch deforms and gets pinned to the Fermi energy. As a
result, a region with compressible densities is formed in the center of the
ribbon. We define the compressible strip as a region where the dispersion
lies within the energy window $|E-E_{F}|\lesssim 2k_{B}T$.\cite%
{Shylau_Magnet} The compressible strips form because in the above energy
window the states are partially filled, i.e. $0\leq f_{FD}\leq 1$ and hence
the system has a metallic character. The formation of compressible strips in
graphene has been recently discussed in \cite{Silvestrov,Shylau_Magnet}.
Further increase of the gate voltage leads to increase of the total electron
density, which, in turn, causes even stronger pinning of the lowest LL to
the Fermi energy such that the compressible strip gets extended almost over
the entire ribbon, see Fig. \ref{fig:BS} (f). Figures \ref{fig:BS} (a)-(c)
show a density distribution for spin-up and spin-down electrons, $%
n_{\uparrow }(y),n_{\downarrow }(y)$ and the total density distribution, $%
n(y)=n_{\uparrow }(y)+n_{\downarrow }(y).$ Due to the strong Zeeman
splitting the electron branch is predominantly spin-up polarized, whereas
the hole branch is spin-down polarized. The total charge distribution shows
a density enhancement towards ribbon's edges. This feature is related to
the effect of the electrostatic Coulomb repulsion in a structure with a
hard-wall confinement and was discussed (for the spinless case) in \cite%
{Silvestrov,Shylau_Cap}.

Compressible strips in GNRs are characterized by the existence of
counterpropagating states at the same edge\cite{Shylau_Magnet}. This
can play an important role in understanding of the magnetotransport
of interacting electrons in the vicinity of the Dirac point. In
one-electron picture the number of propagating states in the region
$-1<\nu <1$ always equals to two with electrons of different spins
propagating near different boundaries, see Fig. \ref{fig:BS} (d).
(Note that a number of propagating
states at a given energy is given by a number of intersections of the line $%
E=E_{F}$ with the dispersion bands $E=E(k),$ and the direction of
the propagation is determined by a sign of $\partial E/\partial
k_{x}$.) The backscattering is probable only in events accompanied
by a spin-flipping since the counterpropagating states with the same
spin are located at the different edges. However in graphene a
spin-relaxation length is rather large and can reach several
micrometers\cite{Tombros}, therefore backscattering is expected to
be rather unimportant. However, the situation is different when
electron-electron interaction is taken into account. In this case
there are a number of additional transport channels at the Fermi
energy where electrons of the same spin propagate in the opposite
directions at the same boundary, see Fig. \ref{fig:BS} (f). In this
case no spin-flipping is apparently needed for the backscattering
and the latter can be caused by e.g. impurity scattering of
electrons of the same spin. To summarize, our self-consistent
calculations demonstrate that, in comparison to the one-electron
picture, electron-electron interaction leads to the drastic changes
in the dispersion relation and structure of propagating states in
the regime of the lowest LL such as a formation of the compressible
strip and opening of additional conductive channels in the middle of
the ribbon.

\begin{figure}[tbh]
\includegraphics[keepaspectratio,width=0.8\columnwidth]{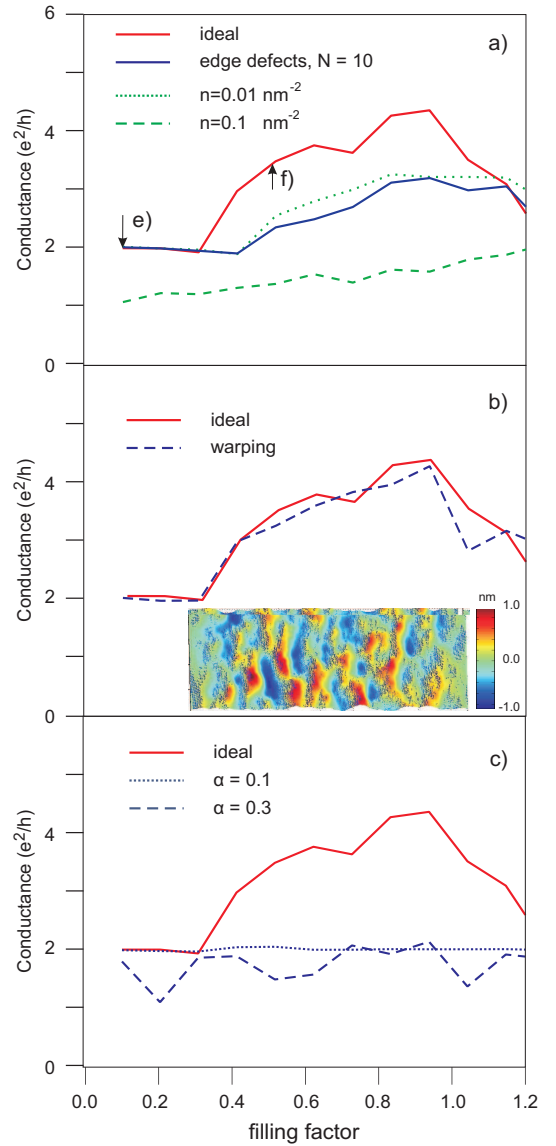}
\caption{(Color online). Conductance of the GNR as a function of the filling
factor $\protect\nu $ calculated in the presence of different types of
disorder: (a) short-range impurities with strength $\protect\delta =4t_{0},$
(b) fluctuation of the normal to the surface component of the magnetic field
due to the warping; (c) spin-flip effect. The red (solid) line on all plots
corresponds to an ideal ribbon (i.e. without disorder). The arrows (e) and
(f) indicate the filling factors for which the corresponding band structures
are plotted on Fig.(\protect\ref{fig:BS}) (e) and (f), respectively. The
inset in (b) illustrates the landscape of a corrugated GNR. A typical
wavelength of the ripples is about 8 nm, while the height fluctuates within
1 nm. $T=4.2$ K.}
\label{fig:Gwarp}
\end{figure}

We proceed to the discussion of the conductance of GNRs in the
regime of the lowest LL taking into account the electron interaction
and spin effects. We will study the effect of different types of
disorder on GNR's conductance, focussing on the robustness of
respectively edge and bulk state transmission. Let us start with the
case of an ideal GNR (i.e. without any disorder) whose conductance
is shown in Fig. \ref{fig:Gwarp} as a red (solid) curve. In the
range $0<\nu \lesssim 0.3$ there are only two states in the energy
interval $4\pi k_{B}T$ and the transport is determined by the edge
states only, see Fig. \ref{fig:BS}(e) (note that in this range of
$\nu $ the states in the compressible regions practically do not
contribute to the
conductance). As a result, the conductance of the ideal ribbon equals to $%
2G_{0}$. Further increase of the filling factor leads to the
enhancement of the conductance due to the additional states forming
the compressible strip, see Fig. \ref{fig:BS}(f). The conductance of
the ideal ribbon is therefore not quantized and develops a bump-like
structure with the increase of the filling factor above $\nu \gtrsim
0.3$. The suppression of conductance quantization in an ideal GNR in
a perpendicular magnetic field for $\nu >1$ was discussed in details
in \cite{Shylau_Magnet}.

Let us now study the effect of impurities on the conductance, see Fig. \ref%
{fig:Gwarp} (a). Recent two-terminal measurements show that the quantized
conductance plateaux are not observed even in high magnetic field\cite%
{Ribeiro,Oostinga,Poumirol}. In particular, this behavior is attributed to
the effect of impurities which can crosslink the chiral currents flowing at
opposite edges.\cite{Ribeiro} We utilize the model of short-range impurities
which are uniformly distributed over the GNR. They are modeled by adding to
the self-consistent potential a term which is randomly chosen in the energy
interval $[-\delta ,\delta ]$, where $\delta =4t_{0}$ is an impurity
strength. The concentration of impurities is chosen in the range $(0.01-0.1)$%
nm$^{-2}$. As expected, the states forming the compressible strip are the
most sensitive to disorder. They propagate in the center of the ribbon and
have a significant spatial extension in the transverse direction. When the
concentration of impurities is moderate (green dotted and blue solid lines),
the conductance decreases but remains larger than $2G_{0}$ since there are
at least two edge states which transmit the current. The edge states are
very robust and can be suppressed only if the impurity concentration greatly
exceeds typical experimental values (green dashed line).

Another source of disorder can arise from a surface warping which is an
inherent property of graphene\cite{Meyer}. Even though the magnetic field is
uniform, a normal to the surface component of the magnetic field fluctuates
due to the corrugated geometry of graphene. It results in a
spatial-dependent hopping integrals. It has been shown that these spatial
correlations can affect the quantum Hall (QH) transition at $E=0$ leading to
anomalously abrupt behaviour\cite{Kawarabayashi} or result in a field-driven
topological transition from the QH-metal state to the QH-insulator state in
the vicinity of $\nu \simeq 0$.\cite{Zhu} In order to take this effect into
account we first generate a corrugated surface (see inset in Fig. \ref%
{fig:Gwarp} (b)) using a method described in Refs. \cite{Fasolino,Klos}.
Then graphene lattice is projected on the surface and a hopping integrals
between two adjacent carbon atoms are calculated by integration of the
vector potential along a line linking these atoms according to Eq. (\ref{phi}%
). We find that the conductance of the warped GNR is practically the same as that one of
the ideal ribbon, see Fig. \ref{fig:Gwarp} (b). We therefore conclude that
for realistic nanoribbons the spatial modulation of the magnetic field due
to the warping has a negligible effect on the conductance.

Finally we investigate the effect of spin-flipping on the
magnetotransport of GNRs. It is modeled by introducing in the Hamiltonian an
additional term $H_{sf}=\alpha \sum_{\mathbf{r}}\left( a_{\mathbf{r}\uparrow
}^{+}a_{\mathbf{r}\downarrow }+a_{\mathbf{r}\downarrow }^{+}a_{\mathbf{r}%
\uparrow }\right) $, where $\alpha $ describes the local spin-flip rate.
This term will in generally admix the counter-propagating states of different
spins leading to backscattering. In our calculations the value of $\alpha $
is varied in the range $(0.0-0.3)t_{0}$\cite{Qiao}. The conductance of GNRs
in the presence of spin-flipping is shown in Fig. \ref{fig:Gwarp}(c). When $%
\alpha =0.1t_{0}$ (dotted line) the states constituting the
compressible strips get scattered and the total conductance is only
due to the edge states. As $\alpha $ increased up to $0.3t_{0}$
(dashed line) the conductance decreases below the value of $2G_{0}$
due to stronger scattering but still remains mostly dominated by the
edge states. Backscattering of spin-polarized states due to
spin-flipping was suggested as the way to manipulate spin
currents\cite{Abanin}. However our results show that even in the
case of strong admixture of counterpropagating states,  the edge
state transmission remains practically unaffected.

\section{Conclusions}

The magnetotransport properties of GNRs in the regime of the lowest LL was
studied taking into account electron interaction, gate electrostatic and
spin effects. When the gate voltage is applied extra charges are induced in
the ribbon leading to an asymmetry between the electron and hole branches in
the energy spectrum. The lowest LL gets pinned to the Fermi energy and a
compressible strip is formed in the middle of the GNR. There are two types
of current carrying states in the lowest LL, the conventional edge states
and the bulk states in the middle of the ribbon constituting the
compressible strip. The later are very sensitive to disorder and get
scattered even if the concentration of disorder is moderate. In contrast,
the edge states are very robust and can not be suppressed even in the
presence of a strong spin-flipping.

\begin{acknowledgments}
We acknowledge a support of the Swedish
Research Council (VR).
\end{acknowledgments}

\end{document}